# On some predictions of the poly-tRNA model for the origin and evolution of genetic coding

By Jacques H. Daniel (1) (2)

((1) Centre National de la Recherche Scientifique, Centre de Génétique Moléculaire, Gif-sur-Yvette, France, (2) InvenTsion, Rehovot, Israel, as present address ;

Correspondence: jacques.daniel1@gmail.com)

The poly-tRNA model was recently presented for the origin and evolution of genetic coding. This model has led to a rather precise description of what might have occurred at the beginning of protein synthesis in the first life form. Here, we further discuss some interesting implications of this model. First, the system of encoded peptide/protein synthesis appears to have started and developed on the breeding ground of a rich RNA world, responsible for the infancy of life existence and complexity. Furthermore, once protein synthesis was fully established and apparently superseding the RNA world, and at a very early stage of life beginnings, we already see what has been a recurrent theme in the likely interpretation of modern comparative molecular studies on species: the full ability of this nascent life entity to develop itself by tinkering, using all kinds of available pieces to improve itself. Lastly, and very instructively, it is deduced from this model that the first peptides to be produced probably had unique properties, not shared by the arsenal of molecules present hitherto, which allowed a functional connection between the then omnipotent RNA world and the lipid membrane vesicles containing it. These specific functions might have initially been at the origin of the Darwinian selection of the full-blown protein-synthesis machinery.

## Introduction

A model was recently proposed for the origin and evolution of genetic coding based on experimental and theoretical arguments (Daniel 2019). In a nutshell, this model contends that the first real-coding genetic setup was a *direct* one, made up of continuous poly-tRNA-like molecules, with each tRNA-like moiety carrying, beyond and near its 5' or 3' end, a trinucleotide site for specific amino acid binding: the sequence and continuity of the tRNA moieties of a particular poly-tRNA would ensure the sequence and continuity of the amino acids of the corresponding peptide or small protein. Then, with the development of a proto-ribosome and a primitive amino acid-activation system, a critical innovation would occur with the appearance of RNA fragments that could tighten several adjacent tRNA moieties together on a particular poly-tRNA molecule, by pairing with the second trinucleotide sequence (identical to the first one carrying the specific amino acid-binding site) situated at, or close to, the middle of each tRNA moiety (i.e., the present "anticodon"site). These fragments, acting as authentic co-ribozymes in the peptide-synthesizing apparatus, would constitute the ancestors of the present mRNAs. Later, on these mRNA-like guiding fragments, free tRNA forms would be additionally used, first keeping their amino acid-binding sites, then losing them in favor of a specific amino acid attachment at a –CCA arm at their 3' end. Finally, these latter mechanisms would progressively prevail, leading to the modern and universal *indirect* genetic coding system (see figure 1).

Experimental results that were in agreement with this poly-tRNA model led to a rather precise description of what might have occurred at the beginning of protein synthesis in the first life form (Daniel 2019). Several different predictions resulting from this description are discussed here.

## Hairpin RNAs and the primitive life form

By comparing the sequences of all tRNAs encoded by two large clusters from *Bacillus subtilis*, a rather robust map could be drawn of the relationships among these various tRNAS (figure 2). For each amino acid tRNA, this map defines the degree of "kinship" with the other tRNAs, as well as the relative time when the particular amino acid entered the protein-synthesis machinery at the dawn of life. Remarkably −barring a few exceptions that are discussed below− this map presents strong regularities that could not be the result of happenstance. Indeed, the amino acids first used for tRNA-based protein synthesis were those considered to have been present naturally in the primordial soup. Moreover, when looking at the particular classes of amino acid-tRNA synthetases (class I or II) to which these amino acids belong, the first amino acids utilized for protein synthesis according to this map were activated by the class II synthetases whereas the later ones were activated by those of class I.

The possible significance of these regularities has been previously discussed. Here, we wish to relate to the other intriguing finding of this study: no apparent regularity between degree of "kinship" and any specific chemical feature of the amino acids. This fact is challenging for scientists, who are accustomed to macromolecular sequence-relationship maps being related to some functional and chemical maps. Since we are concerned here with the construction of the basic bricks of life formation rather than phylogenetic studies at a much later evolutionary time, it was proposed (in view of the poly-tRNA model) that at the origin of the tRNAs there would be one primitive hairpin sequence that could replicate and evolve with time, giving rise to various progeny sequences that would capture, at different evolutionary times, trinucleotide sequences having specific amino acid-binding properties; ligation of two such (probably) identical sequences would generate the primitive tRNA entity particular to each amino acid (see figure 3). These circumstances would account for the apparent fact that there is no co-evolution of the "anticodon" and the rest of the tRNA-generating sequence, but only an unpredictable temporal encounter between the two, thus explaining the quasi-absence of correlation between tRNA sequences and amino acid types or "anticodons".

Therefore, in our model, an essential role is given to hairpin RNAs for gradually creating the various specific tRNAs necessary for the build-up of the robust apparatus for protein synthesis at the heart of every living system today. Is it possible to roughly estimate the number of different hairpin structures that were used for the "purpose" of generating the primordial life form? To arrive at all of the tRNAs for the protein-synthesis machinery, at least one dozen−and more likely several dozen−hairpins may have been necessary. Obviously, it cannot be logically contended that these hairpins were around just for the future encounter with new trinucleotide sequences to make new tRNAs for protein synthesis; therefore, they may have been much more numerous, being certainly involved in other processes. Moreover, these rough number estimates of hairpins might represent only the tip of the iceberg because they may not have had the form (length and perhaps sequence) of those potentially capable of forming tRNAs. So, we are left with even more possibilities of functional and structural hairpin RNAs. And indeed, Dick and Schamel (1995), for example, proposed an RNA-world model that “includes a general stereochemical principle for the interaction between ribozymes and hairpin-derived recognition structures, which can be applied to such seemingly different processes as RNA polymerization, aminoacylation, tRNA decoding, and peptidyl transfer, implicating a common origin for these fundamental functions” ; these authors added that “generation and evolution of tRNA were coupled to the evolution of synthetases, ribosomal RNAs [see also Bloch et al. 1984, 1989; Farias et al. 2014, 2019; Root-Bernstein and Root-Bernstein 2015], and introns from the beginning and have been a consequence arising from the original function of tRNA precursor hairpins as replication and recombination control elements”.

Remarkably, as a recall of its possible crucial role at the origin of life, in the living world today, the RNA hairpin is an essential secondary RNA structure. As mentioned by Svoboda and Di Cara (2006) “it can guide RNA folding, determine interactions in a ribozyme, protect messenger RNA from degradation, serve as a recognition motif for RNA binding proteins or act as a substrate for enzymatic reactions” and be linked to RNA-silencing pathways. As to the tRNA structure, it is known to prime reverse transcription in retroviruses and retrotransposons (Marquet et al. 1995).

**On the relation of the two classes of amino acid-tRNA synthetases to the tRNA map**

As mentioned above, one striking regularity displayed by the tRNA map is that the first amino acids utilized for protein synthesis are activated by the class II synthetases, whereas the later ones are activated by those of class I (see figure 2). However, there are a few exceptions.

The main exceptions concern the tRNAs for valine, leucine and isoleucine which although all belonging to the first group of amino acid tRNAs to enter protein synthesis, are nevertheless recognized and activated by class I synthetases. For valine tRNA, we previously hypothesized that it might have originally belonged to class II because all five tRNAs derived directly from it (glycine, phenylalanine, proline, alanine, and lysine) as well as its only “brother”, tyrosine, belong to class II. Valine, leucine and isoleucine are all aliphatic hydrophobic amino acids, and we can extend this hypothesis of synthetase class shift originally made for valine to both leucine and isoleucine.

Indeed, as another possible argument in favor of the synthetase-class switch having occurred for leucine tRNA , we previously proposed that there might exist some remnants of the primitive genetic coding system, at very precise positions, in the nucleotide sequences that are interspersed between the tRNA genes comprising the two clusters of *Bacillus subtilis*. For class II synthetase tRNA genes, we have found that there exists, at a specific trinucleotide region after the 3’ end of the genes, an excess of nucleotides identical to, and at the same position as, the corresponding anticodons, relative to a specific region before their 5’ end. Moreover, for one threonine tRNA gene, the sequence of the complete anticodon is present after the 3’ end at the predicted position (Daniel 2019). For the valine-, isoleucine- and leucine tRNA genes, a small excess of “nucleotide identity” is also observed after the 3’ end relative to the region before the 5’ end (9 versus 6); in addition, following the 3’ end of one of the leucine tRNA genes, a complete anticodon sequence appears at the expected position (not shown). This suggests that leucine tRNA−and possibly also valine tRNA and isoleucine tRNA−was originally activated by class II synthetases.

What might have been the reason for such a shift? These three aliphatic amino acids have relatively similar hydrophobic chains, which class II synthetase proteins might have had difficulty distinguishing, thus originally rendering protein synthesis less reliable. The later advent of class I synthetases, with a very different and more complex structure at, and close to, their active site, would have allowed better differential recognition of the three amino acids, resulting in much more accurate protein translation. In fact, at about 35 Å from the adenylation site of these amino acids there is a domain, called CP1, that is efficiently involved in post-transfer editing, and it was found, in prokaryotes only, that the CP1 domains of leucine-, valine-, and isoleucine-tRNA synthetases are similar, retaining the ancient form during evolution (Huang et al. 2014). As another (and not exclusive) possibility raised by Bilus et al. (2019), the presence of norvaline in the primitive medium would have required a robust discrimination mechanism to prevent the possible incorporation of this unwanted amino acid into proteins.

Whatever the reason(s) for this postulated switch to class I synthetases occurring with the aliphatic hydrophobic amino acid tRNAs, what could the ancestor molecule of these new synthetases have been? Results from multiple sequence alignment performed with class I synthetases from *Bacillus subtilis* suggest a common origin of the tRNA synthetases for the aliphatic hydrophobic amino acids and that for methionine (figure 4).

The last exception to the synthetase-class rule in the tRNA map is histidine tRNA: although histidine appears to have arrived relatively late in the making of proteins, the cognate synthetase of its tRNA is of class II. The reason for this apparent anomaly may be found in the opportunistic and successful trial of some already existing class II synthetases, or their ancestors, used for other amino acids that had already appeared and were being consumed in protein synthesis, in particular some with a pocket that could, over time, efficiently accommodate the histidine entity. A phylogenetic tree constructed with the multiple protein sequence alignment program Clustal Omega using all class II synthetases from *Bacillus subtilis* shows that histidine-tRNA synthetase has the closest similarity with subunit β of glycine-tRNA synthetase (figure 5; sequence alignment shown in figure 6). Intriguingly, in our tRNA map, histidine tRNA is directly “derived” from glycine(TCC) tRNA (figure 2); in addition, the same sequence is found (5’GCGGXXG) at the 5’ origin of all histidine tRNAs and glycine tRNAs (not shown).

**On the very first encoded peptides**

The poly-tRNA model proposes a precise order of entry of amino acids into the coded peptide/protein-synthesis system during evolution (figure 2). The first one is valine, followed “closely” (no rush in those times) by tyrosine. What might have been the primary reasons for promoting valine as the first coded-system amino acid, to which sometime later tyrosine was added?

The various chemicals naturally contained in the primordial soup might have originally allowed the development of primitive life with no requirement for special metabolic pathways, in particular by providing fatty acid components for the first compartmentalization of what probably were the basic ribozyme reactions and RNA replication (Joyce and Szostak 2018). These native fatty acids may have had many advantageous properties, for starters: by producing closed membranes they could maintain the nascent “life” macromolecules concentrated in a small volume, with no leakage to the environment, and also, protected against genetic parasites (Bansho et al. 2016; Matsumura et al. 2016); at the same time, they could allow the easy diffusion of small molecules into and out of these vesicles; fatty acids could be spontaneously incorporated into the membrane of existing vesicles−which naturally form strong structures−, this vesicle growth being osmotically driven and in competition with other vesicles, i.e., empty vesicles and vesicles encapsulating less efficient systems would shrink by losing membrane (Chen et al. 2004; see below); finally, the fatty acid vesicles could be divided into smaller units as a result of thermal fluctuations, or small mechanical or chemical perturbations, once pearling instability developed (Chen 2009). Szostak's group showed that a hydrophobic dipeptide, $AcPheLeuNH_2$, binds to vesicle membranes, imparting enhanced affinity for fatty acids, and thus promoting vesicle growth (Adamala and Szostak 2013) (this effect was also observed−although with somewhat lesser intensity−with other hydrophobic dipeptides such as PhePhe and $PheLeuNH_2$). This is an interesting finding made in an artificial protocell model, in which $AcPheLeuNH_2$ is catalytically produced inside vesicles by a SerHis dipeptide. However, as the authors state, "Since the catalyst we used is not heritable, our system cannot yet evolve" and in their conclusion, "If such a system exhibited heredity, for example, via the activity of a self-replicating, peptide bond-forming ribozyme, it would amount to a fully functioning protocell, capable of Darwinian evolution."

Here, getting around this issue, we hypothesize that at the very infancy of the protein-synthesis machinery, a di-or tripeptide made of the hydrophobic amino acid valine would have been selected for binding to protocell membranes, leading to enhanced affinity for fatty acids, and thus promoting membrane growth (figure 7). Because these peptides are highly insoluble in water, it is likely that their synthesis would occur at, or very close to, the protocell membrane. Because these peptides would be encoded, their essential primal function would be directly heritable and open to Darwinian evolution.

At some stage, a phospholipid synthase ribozyme might have catalyzed the formation of phospholipids, which are present in membranes of all living beings today. Although low levels of phospholipids would have driven protocell membrane growth during competition for single-chain lipids, this addition to fatty acid membranes would have also made it difficult for electrically- charged compounds to freely diffuse across these more complex charged membranes (Budin and Szostak 2011), likely leading to partial starvation of essential chemicals.

Thus, we postulate that the way out of such an "embarrassment" was to develop the synthesis of some surfactants, in particular some amphipathic Val-Tyr peptide(s) that could greatly improve the diffusion of electrically- charged moieties across the new membranes (figure 8). Remarkably, using UV-visible, surface tension, fluorescence and circular dichroism techniques, it was observed that an artificially made Val-Tyr-Val tripeptide indeed has surfactant activity, being able to produce micelles in aqueous solutions (James and Mandal 2011). In addition, as amphipathic peptides known in the living world today, one can mention the large class of small antimicrobial peptides that have a drastic impact on the structure and integrity of membranes (Khandelia et al. 2008). And again, since this amphipathic Val-Tyr peptide(s) would be encoded and produced by the nascent protein-synthesis apparatus, it would be heritable and capable of Darwinian evolution.

If this interpretation is correct, one of the first encoded peptides might have been selected out for somehow "regulating" the permeability of the novel phospholipid-enriched vesicle membrane to charged compounds: the presence inside vesicles of the normally freely diffusing valine and tyrosine entities would attest to the presence in the environment of all of the precursors needed for the system to enter into growth mode, which would then be allowed by the availability of these precursors−charged or uncharged−thanks to the permeability effect of the valine/tyrosine- containing amphipathic peptide(s). Conversely, when valine and tyrosine would become scarce, the peptide synthesis would stop, and the entry (and exit) of charged molecules would be blocked, resulting in blunt growth arrest.

## Concluding remarks

It is intriguing that the poly-tRNA model previously proposed for the origin and evolution of genetic coding might have rather precise consequences for significant events that may have occurred at the origin of life.

One of these events concerns the temporal emergence of tRNAs and their relation to the amino acid-tRNA synthetase class to which they belong. In the tRNA map presented in the poly-tRNA model, the main exceptions to the synthetase-class rule of early- appearing tRNAs are the three aliphatic hydrophobic amino acid-tRNAs, valine-, leucine- and isoleucine-tRNA (figure 2). It is argued here that at the very origin, these tRNAs might have belonged to class II −in contrast to their present day class I status in all life forms. Presumably, this class shift happened in the protocell soon after the development of protein synthesis, probably when independent tRNAs bore the CCA sequence at their 3’ end (see figure 1) (a shift occurring earlier, with two different tRNAs for each amino acid−one carrying the amino acid-binding site at the 5’ end, and the other at the 3’ end−which would be in competition for protein synthesis [see figure 3, as shown for valine], although theoretically possible, seems nevertheless rather unlikely). In any event, tRNAs for these aliphatic amino acids were all already class I in the last universal common ancestor. If our view is correct, the tRNAs for the aliphatic amino acids abandoned a former synthetase technology (class II) in favor of another technology that came later on and was found to be more adequate for these particular tRNAs ( both “Made In Natural Selection” technologies, of course). This opportunistic event would have many parallels with what several decades of comparative molecular studies on living species have showed us: there is a lot of tinkering and redesigning in the shaping of the biological world, what Jacob (1981) called, many years ago, “bricolage” (cobbling- together). What is striking in the case of these tRNAs is that some rather drastic “bricolage” might have already occurred at the very origin of life.

As another implication of our model, at the arrival of encoded peptide/protein synthesis, it appears that the RNA world−on which primitive life was dependent for its genetic and ribozyme capacities−was flamboyant and ever flourishing, as suggested by the great variety of hairpin RNAs that seem to have populated the primordial cell. The following question can therefore be legitimately asked: what was the need for developing and encoding new kinds of large molecules, such as peptides/proteins?

We believe that the possible answer proposed here would have been hard to formulate without our results pointing to a historical order of appearance of the various amino acid entities into peptides/proteins. Indeed, the two first amino acids to be used for this purpose, according to the model, were valine, and then tyrosine. Valine alone could generate hydrophobic peptides, whereas combined with tyrosine, it could be easily arranged into peptides with amphipathic properties. This strongly suggests that the initial “motivation” for the making of peptides by the protocell would have been the creation of a functional bridge between the genetic/ribozyme RNA world and the vesicle membrane containing it. Building this functional connection probably became necessary at a certain stage of protocell evolution, but no existing type of molecules (RNA or others) was able to entirely fulfill such a role, hence the invention of encoded peptides.

If this view is correct, genetically- encoded protein synthesis would not have been aimed, originally, at replacing the quite ineffective ribozymes by better-performing catalysts; rather, this new system would have filled an important niche that was not occupied by the existing types of molecules, thereby achieving some kind of important regulatory function in relation to the vesicle membrane; it is only later that it would replace most, if not all, ribozyme activities.

Interestingly, RNA-made polymers may have been adopted at the origin of life, at least in part, for their capacity to bind many counterions on their backbone–a property that is not shared, for example, by peptide nucleic acids. This general physicochemical feature at the very start of life would have allowed the buildup of osmotic pressure within the protocell in proportion with its vibrant genetic and ribozyme activities; and the more efficient RNA replication, causing faster cell growth, would lead to " the emergence of Darwinian evolution at the cellular level" (Chen et al. 2004). It is proposed here that at a later protocellular stage, another Darwinian selection – this time, seemingly more specific and properly biologically- oriented–would have operated to achieve regulation of the vesicle membrane's growth and permeability, and this would have concerned a novel class of encoded molecules –the peptides– with hydrophobic or amphipathic characteristics. From here−Darwin obliging−a completely new world would start opening up, based on the protein-synthesis machinery with ultimately 20 different amino acids as raw materials, and which, *in fine*, would revolutionize the entire catalytic and structural foundations of life.

**Figure legends**

**Figure 1**. *Proposed model of the origin and evolution of genetic coding*. Five different coding stages before the advent of modern coding are schematically drawn. Double-lined inverted L-shaped objects, in violet and green, represent primitive tRNAs that are either free, or linked together by spacers to form poly-tRNAs. Red and blue bars represent the nucleotide sequence constituting the amino acid-binding site and “the anticodon site”. Oblique unbroken black line represents the RNA co-ribozyme. Broken lines represent spacer sequences between the tRNAs. See Daniel (2019).

**Figure 2**. *Tree arrangement of the sequence similarity of tRNAs from the two clusters in the* Bacillus subtilis *genome*. For a direct link, the length of the vertical line drawn between two tRNAs is proportional to the dissimilarity index, whose value is equal to 100 minus the similarity

index (SI). The tRNAs of class I are in black, those of class II are in red (generally corresponding to the present tRNA status in bacteria). See Daniel (2019).

**Figure 3**. *Proposed model for generating the various tRNAs*. For each tRNA, and at a certain time during evolution, a hairpin RNA would acquire, at one of its ends, a trinucleotide sequence containing an amino acid-binding site (small horizontal bar in blue, with ”+” sign above or below, it); following duplication (by whatever mechanism) of this expanded hairpin (marked “X2”, with arrow pointing right), a primitive tRNA would be produced with a specific amino acid-binding site at one end, and an “anticodon” in the middle of the molecule (see primitive tRNA at first stage of genetic coding evolution in Fig. 1). To simplify the drawing, after the generation of the second tRNA, blue bars and arrows are implicit. At each tRNA position, only one of the two ends of the hairpin would have accommodated the amino acid-binding site, the same end being used for all of the “amino acids“of the same class (exclusion of the other possibility being expressed in the drawing by parentheses around the other hairpin). For tRNAs with a question mark, another possible mechanism of generation might have been at work, either primarily or secondarily. See Daniel (2019) and text.

**Figure 4**. *Sequence comparison of class I amino acid-tRNA synthetases from* Bacillus subtilis*.* Phylogenetic tree was obtained using the multiple protein sequence alignment program Clustal Omega. Amino acid names refer to their corresponding amino acid-tRNA synthetases. See text.

**Figure 5**. *Sequence comparison of class II amino acid-tRNA synthetases from* Bacillus subtilis*.* Phylogenetic tree was obtained using the multiple protein sequence alignment program Clustal Omega. Amino acid names refer to their corresponding amino acid-tRNA synthetases. See text.

**Figure 6**. *Sequence comparison of histidine-tRNA synthetase and glycine-tRNA synthetase from* Bacillus subtilis*.* The last 60 amino acids of the glycine-tRNA synthetase sequence with no equivalent in the histidine-tRNA synthetase, are omitted. See text.

**Figure 7**. *Effects of peptide(s) made up exclusively of valine residues on vesicle membrane growth.* See text.

**Figure 8**. *Effects of peptide(s) made up of valine and tyrosine residues on protocell growth*. See text.

**Figure 1**

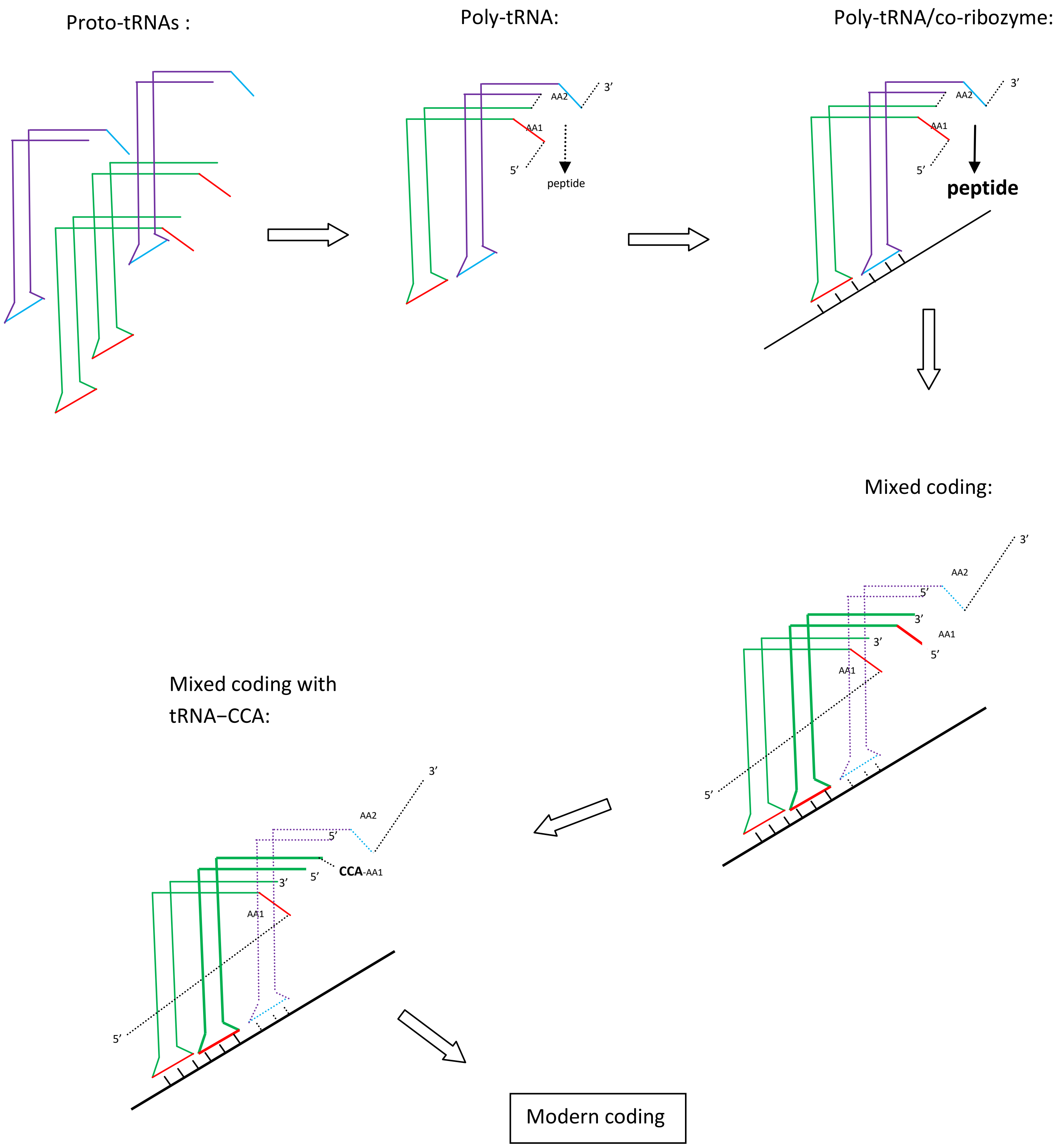

**Figure 2**

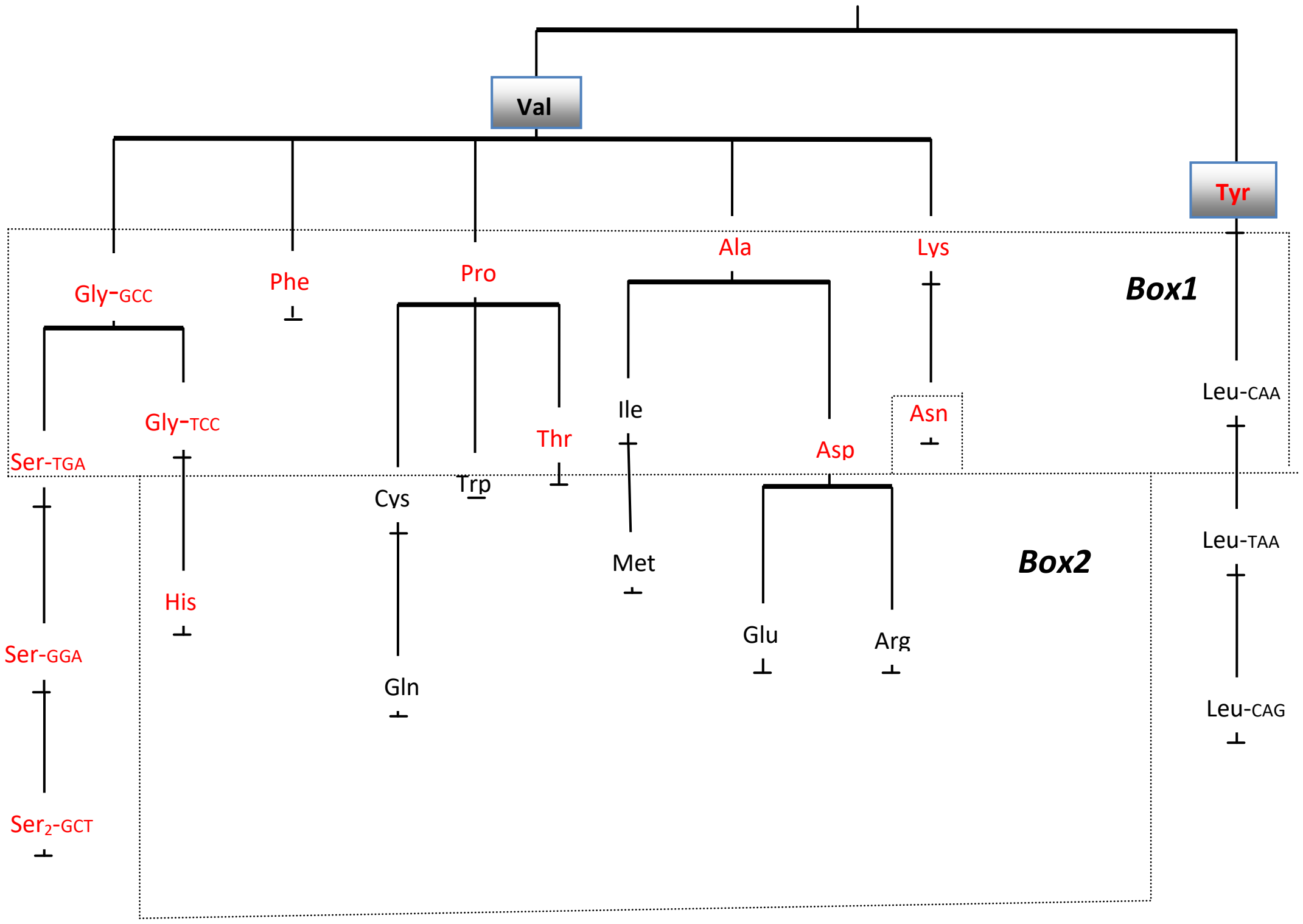

## Figure 3

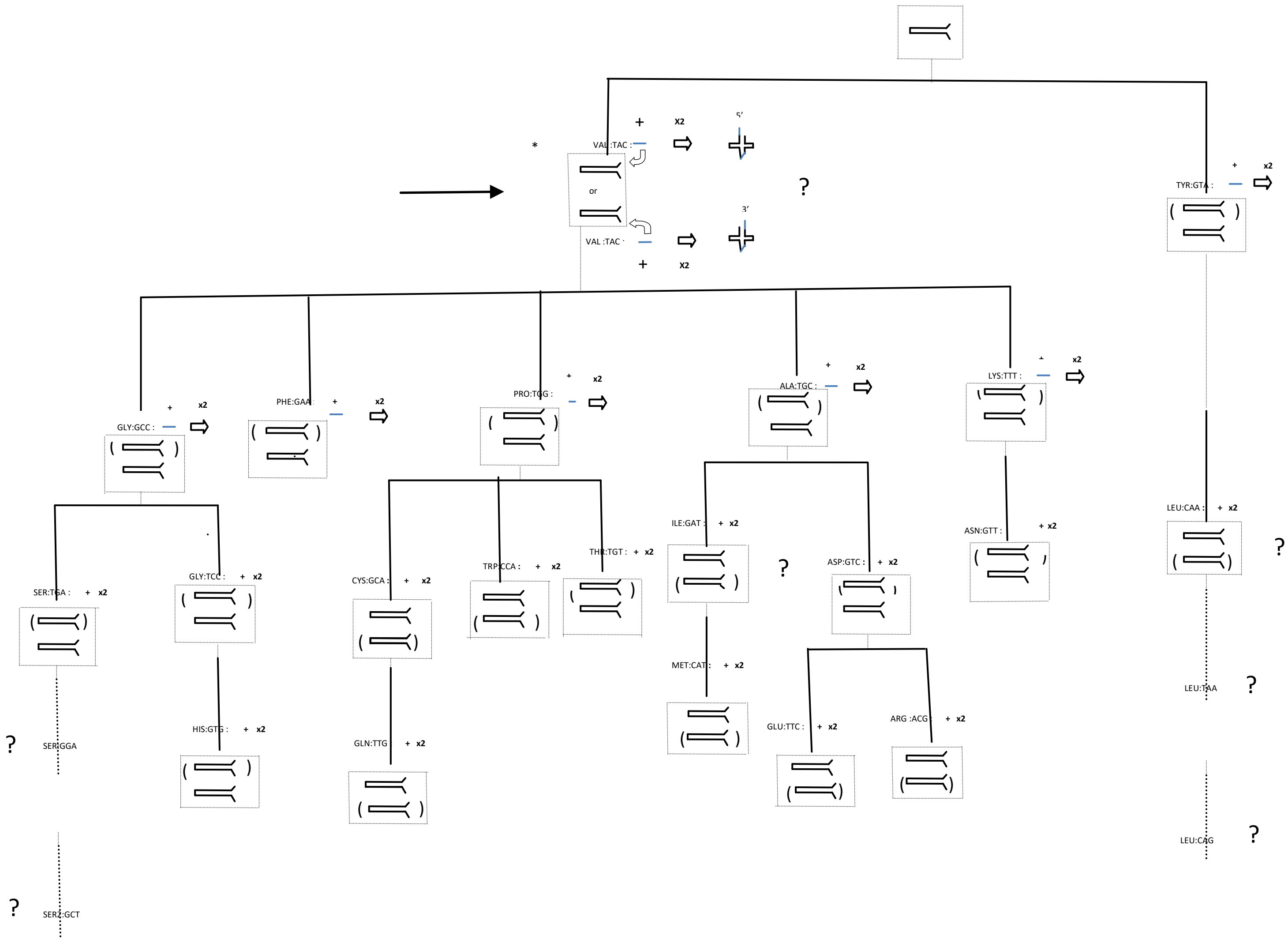

**Figure 4**

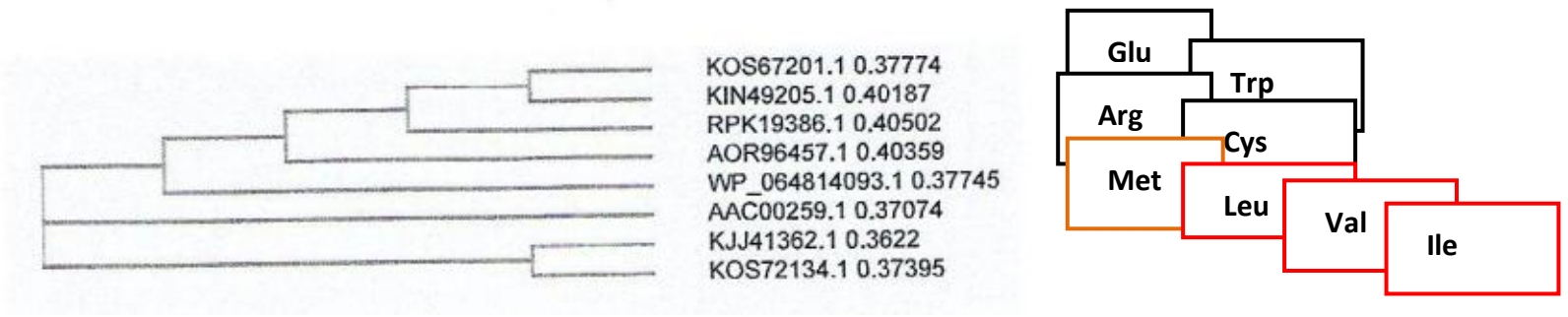

**Figure 5**

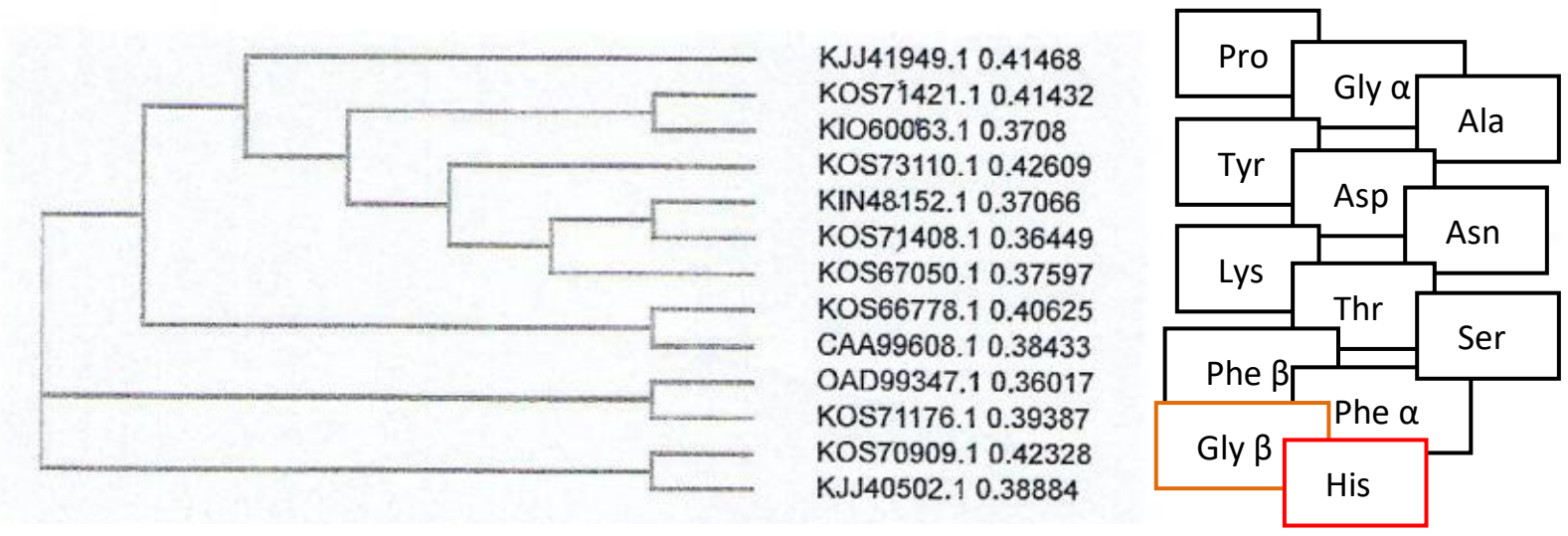

**Figure 6**

```
KJJ40502.1      ---------MGY-----NIPRGTQDILPGESDRWQFVEQIMRDTCRTYQYKE--------    38
KOS70909.1      MSKQDLLLEIGLEEMPARFLNESMVQLGDKLTGWLKEKNITHGEVKLFNTPRRLAVFVKD    60
                         :*      .: . :   * .:   *   ::* :.  : ::  .

KJJ40502.1      -----------IRTPI-----------FEHTELFARGVGESTD-IVQK----EMYTFQDR    71
KOS70909.1      VAEKQDDIKEEAKGPAKKIALDADGNWTKAAIGFSKGQGANVEDLYIKEVKGIEYVFVQK    120
                            : *             : :  *::* * ..: :  *      *.* ::

KJJ40502.1      K--GR-SLTLRPEGTAAAVRAFNENKLFANPVQPTKLYYVGPMFRYERPQTGRFRQF---    125
KOS70909.1      FQAGQETKSLLPELS----------GLITSLHFPKNMRWGNEDLRYIRPIKWIVALFGQD    170
                   *: : :* ** :           *::.   *.:: : .  :** **  .  .  *

KJJ40502.1      -YQFGIEAIGSKDPAIDAEVMALAMSIYQKAGLEN----VKL------------------    162
KOS70909.1      VIPFSITNVESGRTTQGHRFLGHEVSIESPSAYEEQLKERHVIADPSVRKQMIQSQLEAM    230
                   *.*  : *   : . ..:.  :** . :. *:     ::

KJJ40502.1      ----------VINSLGDQESRKSYREALVKHFEPRIEEFCSDCQSRLHTNPLRILDCKKD    212
KOS70909.1      AAENDWSIPVDEDLLDEVNHLVEYPTALYGSFESEFLSIPEEVLVTTMKEHQRYFPV-KD    289
                            : *.: :   .*  **   ** .: .: .:      .:  * :   **

KJJ40502.1      RDHELMKSAPSILTYLNE--------------------ESAAYFEKVKQY----------    242
KOS70909.1      KNGDLL---PHFITVRNGNSHAIENVARGNEKVLRARLSDASFFYKEDQKLNIDANVKKL    346
                :: :*:   * ::*  *                     ..*::* * .*

KJJ40502.1      ----------------------LNDLGISYEIDPNL---------VRGLDYYNHTAFEIM    271
KOS70909.1      ENIVFHEELGSLADKVRRVTSIAEKLAVRLQADEDTLKHVKRAAEISKFDLVTHMIYEFP    406
                                       :.*.:  : * :          :  :*  .*  :*:

KJJ40502.1      S-----------------------------NAE-----------------------GFGA    279
KOS70909.1      ELQGIMGEKYARMLGEDEAVAAAVNEHYMPRSAGGETPSTFTGAVVAMADKLDTIASFFS    466
                .                             .:                        .* :

KJJ40502.1      ITTLA-GGGRYDGLVEQIGGPE--VPGIGFAMSIERLLAAIDAEKRELPVDQGIDCYIVT    336
KOS70909.1      IGVIPTGSQDPYGLRRQASGIVAILLDRNWGISFEELLTFVQTDKEIELL----DF----    518
                * .:  *.    ** .* .*    : . .:.:*:*.**: ::::*.   :    *

KJJ40502.1      LGEKAKDYSVSLVYKLREAGISSEIDYENKKMKGQFKAADRLNARFIAILGEDELAQN--    394
KOS70909.1      -------FTQRLKYVLNAEQIRHD-------------------VIDAVLESSELEPYSA    551
                       ::  * * *.   *  :                     : *:* ..**

KJJ40502.1      --KINVKDAQTG----EQIEVALDEFIHVMKANQKG------------------------    424
KOS70909.1      LHKAQVLEQKLGAPGFKETAEALGRVISISKKGVRGDIQPDLFENEYEAKLFDAYQTAKQ    611
                  * :* : : *    ::   **...* : * . :*

KJJ40502.1      ------------------------------------------------------------    424
```

https://www.ebi.ac.uk/Tools/services/rest/clustalo/result/clustalo-I20200714-110003-0517-32132342-p1m/aln-clustal_num

**Figure 7**

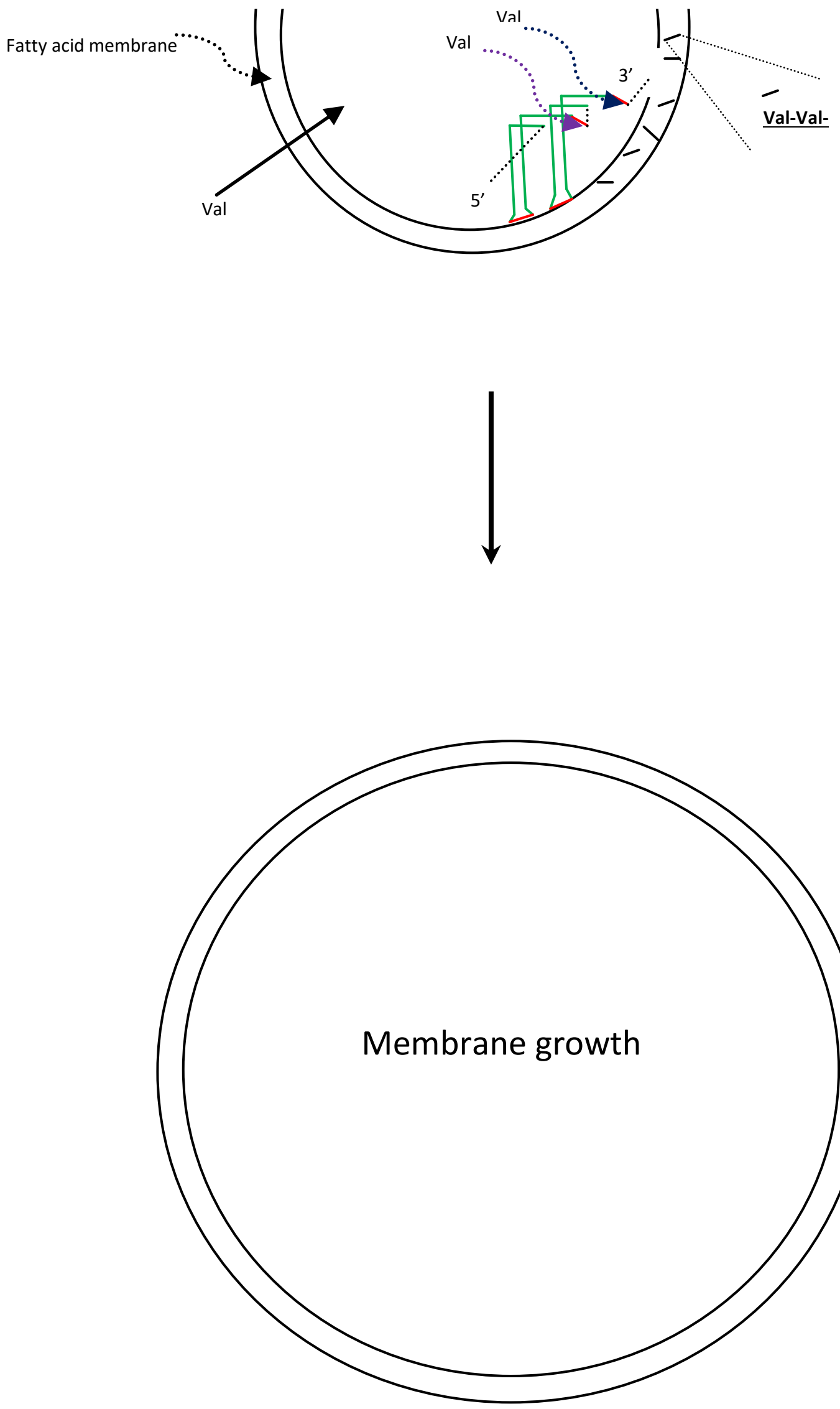

**Figure 8**

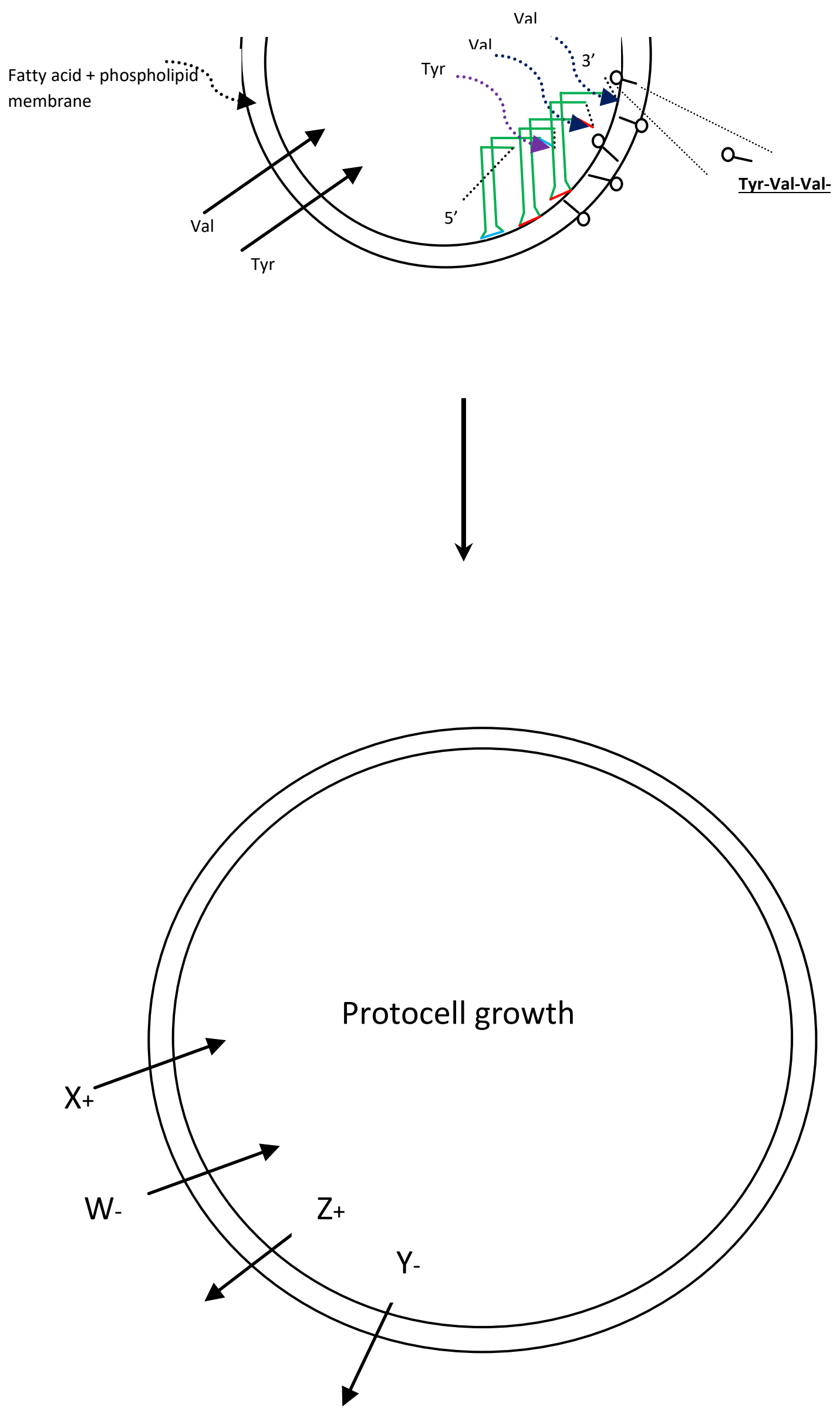